\documentstyle[12pt,subeq]{article}

\newcommand\vek[1]{\mbox{\rmfamily\bfseries\itshape#1}}

\begin{document}

\title{Microscopic Calculation of the Dielectric Susceptibility
Tensor for Coulomb Fluids}

\author{L. \v Samaj \\ Institute of Physics, Slovak Academy of Sciences,
D\' ubravsk\' a cesta 9, \\ 842 28 Bratislava, Slovakia; e-mail:
fyzimaes@savba.sk}

\maketitle

\begin{abstract}
In a Coulomb fluid confined to a domain $V$, the dielectric 
susceptibility tensor $\chi_V$ depends on the shape of $V$, 
even in the thermodynamic $V\to\infty$ limit.
This paper deals with the classical two-dimensional one-component plasma
formulated in an elliptic $V$-domain, 
equilibrium statistical mechanics is used.
For the dimensionless coupling constant $\Gamma$ = even positive integer, 
the mapping of the plasma onto a discrete one-dimensional 
anticommuting-field theory provides a new sum rule.
This sum rule confirms the limiting value of $\chi_V$ predicted by
macroscopic electrostatics and gives the finite-size correction term
to $\chi_V$.
\end{abstract}


\vfill


\newpage

\section{Introduction}

Classical Coulomb systems are the prototypes for studying the effect
of long-range interactions in equilibrium statistical mechanics.
In dimension $\nu$, the Coulomb potential $\phi^c$ at spatial position
${\vek r} = (r^1, r^2,\ldots, r^{\nu})$, induced by a unit charge at
the origin, is the solution of Poisson equation
\begin{equation} \label{1}
\Delta \phi^c({\vek r}) = - \epsilon_{\nu} \delta({\vek r})
\end{equation}
with $\epsilon_2 = 2\pi, \epsilon_3 = 4\pi$, etc.
In particular, in $\nu = 2$ dimensions, one has
\begin{equation} \label{2}
\phi^c({\vek r}) = - \ln (\vert {\vek r} \vert / r_0)
\end{equation}
where $r_0$ is a length scale, for simplicity, set to unity.
A general Coulomb system consists of $s$ pointlike species 
$\alpha = 1,\ldots, s$ of the corresponding charges $q_{\alpha}$,
embedded in a fixed uniform neutralizing background of density $n_0$
and charge density $\rho_0$.
The most studied one-component jellium or plasma (OCP) and the two-component
plasma (TCP) correspond to $s=1 ~ (q_1=q), \rho_0 = -q n_0 \ne 0$ and to 
$s=2 ~ (q_1 = - q_2), \rho_0 = n_0 = 0$, respectively.
The Coulomb system is confined to a domain $V$, which can be:

(1) infinite, $V \to {\bf R}^{\nu}$;

(2) finite, bounded by an impermeable hard wall (for the sake of simplicity, 
uncharged and with no image forces);

(2) semi-infinite, i.e., bounded by a wall, but infinite in at least
one of the parallel directions.

\noindent The symbol $\langle \ldots \rangle_V$ will denote
the canonical averaging over the domain $V$ at the inverse temperature
$\beta = 1/k_B T$, under the system neutrality condition.
The microscopic total number and charge densities at ${\vek r}$
are given by
\begin{subequations} \label{3}
\begin{eqnarray}
\hat n({\vek r}) & = & \sum_i \delta({\vek r} - {\vek r}_i) \\
\hat \rho({\vek r}) & = & \sum_i q_{\alpha_i} 
\delta({\vek r} - {\vek r}_i)
\end{eqnarray}
\end{subequations}
respectively, where the sums run over $N$ particle indices.
The canonical average number and charge densities read
\begin{equation} \label{4}
n_V({\vek r}) = \langle \hat n({\vek r}) \rangle_V, \quad \quad
\rho_V({\vek r}) = \langle \hat \rho({\vek r}) \rangle_V 
\end{equation}
At the two-particle level, one considers
the two-body distribution
\begin{subequations} \label{5} 
\begin{equation} \label{5a} 
n_V({\vek r},{\vek r}') = \langle \sum_{j\ne k} 
\delta({\vek r} - {\vek r}_j) \delta({\vek r}' - {\vek r}_k) \rangle_V 
\end{equation}
as well as its truncated form
\begin{equation} \label{5b}
n_V^T({\vek r},{\vek r}')  =  n_V({\vek r},{\vek r}') 
- n_V({\vek r}) n_V({\vek r}')
\end{equation}
\end{subequations}
and the truncated charge-charge correlation function
\begin{equation} \label{6}
S_V({\vek r} \vert {\vek r}') = 
\langle \hat \rho({\vek r}) \hat \rho({\vek r}') \rangle_V - 
\langle \hat \rho({\vek r}) \rangle_V \langle \hat \rho({\vek r}') \rangle_V 
\end{equation}
In the case of the OCP, $S$ is expressible as follows
\begin{equation} \label{7}
S_V({\vek r}\vert {\vek r}') = q^2 \left[ n_V^T({\vek r},{\vek r}')
+ n_V({\vek r}) \delta({\vek r} - {\vek r}') \right]
\end{equation} 

The long-range tail of the Coulomb force causes screening,
and thus gives rise to exact constraints, sum rules, for 
the structure function $S$ (see review \cite{Martin}).
Their explicit form depends on the geometry of domain $V$.

In the bulk regime, $\lim_{V \to {\bf R}^{\nu}} 
S_V({\vek r}\vert {\vek r}') = S(\vert {\vek r}- {\vek r}' \vert)$ is known 
to obey the Stillinger-Lovett rules \cite{Stillinger1},\cite{Stillinger2} 
which imply the zeroth-moment (electroneutrality) condition
\begin{equation} \label{8}
\int {\rm d}{\vek r} ~ S({\vek r}) = 0
\end{equation}
and the second-moment condition
\begin{eqnarray} \label{9}
\beta \int {\rm d}{\vek r} ~ ( r^i )^2 S({\vek r})
& = & {\beta \over \nu} \int {\rm d}{\vek r} ~ \vert {\vek r} \vert^2 
S({\vek r}) \nonumber \\
& = & -  {1\over \pi (\nu -1)}  \quad \quad i=1,\ldots,\nu
\end{eqnarray}
For the OCP, the fourth moment of $S$ is related to the isothermal
compressibility \cite{Pines},\cite{Vieillefosse1},\cite{Baus}, so that 
the knowledge of the exact equation of state in dimension two 
\cite{Hauge} provides the explicit form of the former
\cite{Vieillefosse2},\cite{Suttorp}.
Very recently \cite{Kalinay}, the sixth moment of $S$ for the 2d OCP
was derived using a renormalized Mayer expansion in density \cite{Deutsch}. 
The formal analogues of the fourth and sixth moments of $S$ in
the 2d OCP are, respectively, the zeroth and second moments of the
truncated total number density correlation function
$\langle \hat n({\vek r}) \hat n({\vek r}')\rangle -
\langle \hat n({\vek r}) \rangle \langle \hat n({\vek r}')\rangle$
in the 2d TCP, as was derived in Ref. \cite{Jancovici1} from analogies
with critical systems and in Ref. \cite{Jancovici2} directly by using
diagrammatic methods. 

For finite systems, the sum rule
\begin{equation} \label{10}
\int_V {\rm d}{\vek r} ~ S_V({\vek r} \vert {\vek r}') =
\int_V {\rm d}{\vek r}' ~ S_V({\vek r} \vert {\vek r}') = 0
\end{equation}
only tells us that the total charge in domain $V$ is fixed.
The information analogous to the second-moment formula (\ref{9})
is contained in the dielectric susceptibility tensor $\chi_V$,
defined by
\begin{subequations} 
\begin{equation} \label{11a}
\chi_V^{ij} = {\beta \over \vert V \vert} \left( \langle P^i P^j \rangle_V
- \langle P^i \rangle_V \langle P^j \rangle_V \right) 
\end{equation}
where
\begin{equation} \label{11b}
P^i = \int_V {\rm d} {\vek r} ~ r^i \hat \rho({\vek r}) \quad \quad
i = 1,\ldots, \nu
\end{equation}
\end{subequations}
is the $i$th component of the total polarization in the system
and $\vert V \vert$ the volume.
Within linear-response theory, $\chi_V$ relates the average polarization
to a constant applied field ${\vek E}$, 
$\langle P^i \rangle = \sum_{j=1}^{\nu} \chi_V^{ij} E^j$.
With regard to (\ref{10}), the tensor $\chi_V$ is expressible in 
two equivalent ways,
\begin{eqnarray} \label{12}
\chi_V^{ij} & = & {\beta \over \vert V \vert } \int_V {\rm d}{\vek r}_1
\int_V {\rm d}{\vek r}_2 ~ r^i_1 r^j_2 S_V({\vek r}_1\vert {\vek r}_2)
\nonumber \\
& = & - {\beta \over 2 \vert V \vert } \int_V {\rm d}{\vek r}_1
\int_V {\rm d}{\vek r}_2 \left(r^i_1 - r^j_2 \right)^2 
S_V({\vek r}_1\vert {\vek r}_2) 
\end{eqnarray}
As $V\to {\bf R}^{\nu}$ one would intuitively expect that,
according to the bulk second-moment formula (\ref{9}), the diagonal 
components $\chi_V^i = \chi_V^{ii}$ $(i=1,\ldots,\nu)$ tend to the value
\begin{equation} \label{13}
\chi_V^i \to - {\beta \over 2} \int {\rm d}{\vek r} ~ ( r^i )^2
S({\vek r}) = {1\over 2\pi (\nu -1)} 
\end{equation}
However, this is not the case: due to surface effects, the $\chi_V$ limit
depends on the shape of $V$.
Its value is predicted by macroscopic electrostatics for homogeneously 
polarizable systems \cite{Choquard1},\cite{Choquard2}.
In the case of elliptic $(\nu = 2)$ and ellipsoidal $(\nu = 3)$ $V$-domains,
one introduces the depolarization tensor $T_V$
\begin{equation} \label{14}
T_V^{ij} =  - {1\over 2\pi (\nu -1)} {\partial^2 \over \partial r^i 
\partial r^j} \int_V {\rm d}{\vek r}' ~ \phi^c({\vek r}-{\vek r}')
\end{equation}
where ${\vek r}$ is an arbitrary point in $V$.
It is the fundamental property of elliptic and ellipsoidal domains that
the tensor $T_V$ is independent of point ${\vek r}\in V$, and depends only 
on the shape of $V$.
With regard to Poisson equation (1), its diagonal elements 
$T_V^i = T_V^{ii}$ are constrained by 
$\sum_{i=1}^{\nu} T_V^i  =  \epsilon_{\nu}/[2\pi(\nu-1)]$. 
In the limit $V\to {\bf R}^{\nu}$, electrostatics yields
\begin{equation} \label{15}
\chi_V^i = {1\over 2\pi (\nu -1) T_V^i}
\end{equation}
In the special case of a 2d disk or a 3d sphere, $T_V$ is isotropic, so that
$T_V^i = \epsilon_{\nu}/[2\pi\nu(\nu-1)]$. Consequently, 
\begin{equation} \label{16}
\chi_V^i \to {\nu \over \epsilon_{\nu}} = {\nu \over 2\pi (\nu -1)} 
\quad \quad {\rm for} \ \nu=2,3
\end{equation}
in contradiction with the previously suggested estimate (\ref{13}).
Formula (\ref{16}) was checked and finite-size corrections were calculated
in \cite{Choquard1},\cite{Choquard2} for the 2d OCP formulated on disk
for the exactly solvable case of the dimensionless coupling 
$\Gamma = \beta q^2 =2$ \cite{Jancovici3}, \cite{Jancovici4}.

This paper deals with the 2d OCP formulated in an elliptic domain, 
which includes the circularly symmetric disk and the limiting case, the strip. 
The statistics now depends on the only parameter -- the coupling
constant $\Gamma$ (the particle density only scales
appropriately the distance).
At $\Gamma$ = even integer, the 2d OCP is mappable onto a discrete
1d anticommuting-field theory \cite{Samaj1},\cite{Samaj2}.
It is shown that sum rules come from specific unitary 
transformations of anticommuting variables, keeping the general form 
of the fermionic action.
A nontrivial transformation of anticommuting variables is revealed
to generate a new sum rule.
For the elliptic domain, this sum rule confirms microscopically 
the asymptotic formula (\ref{15}) and gives the finite-size
correction term to $\chi^i_V$ explicitly in terms of boundary
contributions.

The paper is organized as follows.
Section 2 recapitulates briefly the mapping of the 2d OCP
onto the 1d fermionic model.
Section 3 establishes the formalism of unitary transformations of 
anticommuting variables, which imply the known sum rules \cite{Tellez}.
Complementary (to the author's knowledge as-yet-unknown) sum rules, 
nontrivial when some asymmetry of the $V$-domain 
is present, are given as well. 
In the key Section 4, using a special ``nearest-neighbor''
transformation of anticommuting variables, one derives a new sum rule 
providing a proper split of $\chi_V^i$ into its asymptotic (\ref{15})
and finite-size correction parts. 
In Appendix, by explicit calculations in the 2d OCP on disk at $\Gamma=2$, 
a test of the results is presented.

\section{Mapping onto 1d fermions}

The model under consideration is the 2d OCP of $N$ particles confined 
to a domain $V$.
For point ${\vek r} \in V$, the cartesian $(x,y)$, complex 
$(z,\bar z)$ or polar $(r,\phi)$ coordinate representations will be 
suitably used.
The neutralizing background of density $n_0 = N/\vert V\vert$
induces the one-particle potential
\begin{equation} \label{17}
\phi_b({\vek r}) = \int_V {\rm d}^2 r' ~ \phi^c(\vert {\vek r}
- {\vek r}' \vert)
\end{equation}
For the elliptic $V$-domain of interest, in the reference frame defined
by the axis of the ellipse, $x^2/a^2+y^2/b^2=1$, both tensors $\chi_V$
and $T_V$ are diagonal. 
The fundamental independence of the depolarization tensor $T_V$ (\ref{14}) 
of point ${\vek r} \in V$ and the invariance of $\phi_b({\vek r})$ 
with respect to the reflection along the $x$ or $y$ axis imply
\begin{subequations} 
\begin{equation} \label{18a}
\phi_b({\vek r}) = {\rm const} - \pi T_V^x x^2 - \pi T_V^y y^2
\end{equation}
with $T_V^x = b/(a+b), T_V^y = a/(a+b)$.
The corresponding electric field reads
\begin{equation} \label{18b}
{\vek E}_b({\vek r}) = - \nabla \phi_b({\vek r})
= 2 \pi T_V^x x {\hat{\vek x}} + 2 \pi T_V^y y {\hat{\vek y}}
\end{equation}
\end{subequations}
where ${\hat{\vek x}},{\hat{\vek y}}$ are the perpendicular unit vectors 
in $x,y$ directions.
In the circularly symmetric case of disk $a=b=R$ (radius), one has
$T_V^x = T_V^y = 1/2$, so that
\begin{equation} \label{19}
\phi_b({\vek r}) = - \pi r^2 / 2 , \quad \quad 
{\vek E}_b({\vek r}) = \pi {\vek r}
\end{equation}
The total Boltzmann factor associated with particle configuration
$\{ {\vek r}_i \}$ is written as
\begin{equation} \label{20}
\exp \big[ \Gamma n_0 \sum_i \phi_b({\vek r}_i) - \Gamma \sum_{i<j}
\phi^c(\vert {\vek r}_i - {\vek r}_j \vert) \big]
\end{equation}

For $\Gamma = 2 \gamma$, $\gamma$ being positive integer, it was
shown in \cite{Samaj1} that the canonical partition function of
the system, $Z_V$ (we will omit in the notation the dependence on $N$),
can be expressed in terms of Grassmann variables
$\{ \xi_i^{(\alpha)}, \psi_i^{(\alpha)} \}$ $(\alpha = 1,\ldots,\gamma)$,
defined on sites $i = 0, 1,\ldots, N-1$ of a discrete chain and
satisfying the ordinary anticommuting algebra and integration rules
\cite{Berezin}, as follows:
\begin{subequations} \label{21}
\begin{eqnarray}
Z_V & = & \int {\cal D}\psi {\cal D}\xi ~ \exp[{\cal S}_V(\xi,\psi)] \\
{\cal S}_V(\xi,\psi) & = & \sum_{i,j=0}^{\gamma (N-1)} \Xi_i w_{ij} \Psi_j
\end{eqnarray}
\end{subequations}
Here, ${\cal D}\psi {\cal D}\xi = \prod_{i=0}^{N-1} {\rm d}\psi_i^{(\gamma)}
\ldots {\rm d}\psi_i^{(1)} {\rm d}\xi_i^{(\gamma)}\ldots {\rm d}\xi_i^{(1)}$
and the fermionic action ${\cal S}_V$ involves pair interactions 
of ``composite'' variables
\begin{subequations} \label{22}
\begin{eqnarray}
\Xi_i & = & \sum_{i_1\ldots i_{\gamma}=0 
\atop (i_1+\ldots+i_{\gamma}=i)}^{N-1}
\xi_{i_1}^{(1)} \ldots \xi_{i_{\gamma}}^{(\gamma)} \label{22a} \\
\Psi_j & = & \sum_{j_1\ldots j_{\gamma}=0 
\atop (j_1+\ldots+j_{\gamma}=j)}^{N-1}
\psi_{j_1}^{(1)} \ldots \psi_{j_{\gamma}}^{(\gamma)} \label{22b}
\end{eqnarray}
\end{subequations}
i.e., the products of $\gamma$ anticommuting-field components
with a given sum of site indices.
The interaction strength is given by
\begin{equation} \label{23}
w_{ij} = \int_V {\rm d}^2 z ~ z^i {\bar z}^j w(z,{\bar z})
\end{equation}
where $w({\vek r}) = \exp[\Gamma n_0 \phi_b({\vek r})].$
Denoting by
\begin{equation} \label{24}
\langle \ldots \rangle = {1\over Z_V} \int {\cal D}\psi
{\cal D}\xi ~ \ldots \exp[ {\cal S}_V(\xi,\psi)]
\end{equation}
the averaging over the $1d$ fermionic system, the particle-number
density (\ref{4}) can be expressed as
\begin{equation} \label{25}
n_V({\vek r}) = w(z,{\bar z}) \sum_{i,j=0}^{\gamma(N-1)}
\langle \Xi_i \Psi_j \rangle ~ z^i {\bar z}^j
\end{equation}
the two-body distribution (\ref{5a}) and its truncation (\ref{5b}) as
\begin{subequations} \label{26}
\begin{eqnarray}
n_V({\vek r}_1,{\vek r}_2) & =  & w(z_1,{\bar z}_1) w(z_2,{\bar z}_2) 
\sum_{i_1 j_1 i_2 j_2 =0}^{\gamma(N-1)}
\langle \Xi_{i_1} \Psi_{j_1} \Xi_{i_2} \Psi_{j_2} \rangle ~ 
z_1^{i_1} {\bar z}_1^{j_1} z_2^{i_2} {\bar z}_2^{j_2} \\
n_V^T({\vek r}_1,{\vek r}_2) & =  & w(z_1,{\bar z}_1) w(z_2,{\bar z}_2) 
\sum_{i_1 j_1 i_2 j_2 =0}^{\gamma(N-1)}
\langle \Xi_{i_1} \Psi_{j_1} \Xi_{i_2} \Psi_{j_2} \rangle^T ~ 
z_1^{i_1} {\bar z}_1^{j_1} z_2^{i_2} {\bar z}_2^{j_2}
\end{eqnarray}
\end{subequations}
where $\langle \Xi_{i_1} \Psi_{j_1} \Xi_{i_2} \Psi_{j_2} \rangle^T = 
\langle \Xi_{i_1} \Psi_{j_1} \Xi_{i_2} \Psi_{j_2} \rangle -
\langle \Xi_{i_1} \Psi_{j_1} \rangle 
\langle \Xi_{i_2} \Psi_{j_2} \rangle$.
For disk (\ref{19}), since the Boltzmann weight $w({\vek r})$ 
possesses the circular symmetry, the interaction matrix $w_{ij}$ 
takes the diagonal form,
\begin{equation} \label{27}
w_{ij} = \delta_{ij} w_i , \quad \quad  w_i = \int_V {\rm d}^2 r ~ r^{2i} w(r)
\end{equation}
Owing to the ``diagonalization'' of the action,
${\cal S}_V = \sum_i \Xi_i w_i \Psi_i$, only fermionic correlators
$\langle \Xi_{i_1} \Psi_{j_1} \Xi_{i_2} \Psi_{j_2} \ldots \rangle$
with $i_1+i_2+\ldots = j_1+j_2+\ldots$ survive.

\section{Ordinary sum rules and their complements}

Sum rules result from the fermionic representation of the 2d OCP
by specific transformations of anticommuting variables, keeping
the composite nature of the action ${\cal S}_V$ (21b).

Let us first rescale by a constant one of the field components,
say $\{ \xi^{(1)} \}$,
\begin{equation} \label{28}
\xi_i^{(1)} \to \mu \xi_i^{(1)} \quad \quad i = 0, 1, \ldots, N-1
\end{equation}
Jacobian of the transformation equals to $\mu^N$ and the fermionic action
${\cal S}_V$ transforms to $\mu {\cal S}_V$.
Consequently,
\begin{subequations} \label{29}
\begin{eqnarray}
Z_V & = & \mu^{-N} \int {\cal D}\psi {\cal D}\xi ~ 
\exp \big( \mu \sum_{i,j=0}^{\gamma (N-1)} \Xi_i w_{ij} \Psi_j \big) \\
Z_V \langle \Xi_i \Psi_j \rangle & = & \mu^{-N+1} 
\int {\cal D}\psi {\cal D}\xi ~ \Xi_i \Psi_j
\exp \big( \mu \sum_{k,l=0}^{\gamma (N-1)} \Xi_k w_{kl} \Psi_l \big)
\end{eqnarray}
\end{subequations}
etc.
$Z_V$, a Grassmanian integral, is independent of $\mu$, thus its
derivative with respect to $\mu$ is zero for any value of $\mu$.
For the special case $\mu =1$, the equality 
$\partial \ln Z_V / \partial \mu \vert_{\mu=1} = 0$ implies
\begin{equation} \label{30}
-N + \sum_{i,j=0}^{\gamma(N-1)} w_{ij} \langle \Xi_i \Psi_j \rangle = 0
\end{equation}
which, after substituting (\ref{23}), regarding (\ref{25}) and setting
$N = n_0 \vert V \vert$, results in the trivial neutrality condition
\begin{equation} \label{31}
\int_V {\rm d}^2 r ~ \left[ n_V({\vek r}) - n_0 \right] = 0
\end{equation}
Analogously, the equality $\partial [ Z_V \langle \Xi_i \Psi_j \rangle ] / 
\partial \mu \vert_{\mu=1} = 0$ yields
\begin{equation} \label{32}
- (N-1) \langle \Xi_i \Psi_j \rangle + \sum_{k,l=0}^{\gamma(N-1)}
w_{kl} \langle \Xi_i \Psi_j \Xi_k \Psi_l \rangle = 0
\end{equation}
which is readily shown to be equivalent to the neutrality relation (\ref{10}).

Let us now consider another linear transformation of {\it all} $\xi$-field
components,
\begin{subequations} 
\begin{equation} \label{33a}
\xi_i^{(\alpha)} \to \lambda^i \xi_i^{(\alpha)} \quad \quad
i=0,1,\ldots,N-1; \quad \alpha = 1,\ldots,\gamma
\end{equation}
or {\it all} $\psi$-field components,
\begin{equation} \label{33b}
\psi_j^{(\alpha)} \to \lambda^j \psi_j^{(\alpha)} \quad \quad
i=0,1,\ldots,N-1; \quad \alpha = 1,\ldots,\gamma
\end{equation}
\end{subequations}
Jacobian of both transformations equals to $\lambda^{\gamma N(N-1)/2}$
and the action transforms as 
${\cal S}_V \to \sum_{i,j=0}^{\gamma(N-1)} \lambda^i \Xi_i w_{ij} \Psi_j$
under (\ref{33a}) and as
${\cal S}_V \to \sum_{i,j=0}^{\gamma(N-1)} \lambda^j \Xi_i w_{ij} \Psi_j$
under (\ref{33b}).
Thus,
\begin{subequations} \label{34}
\begin{eqnarray}
Z_V & = & \lambda^{-\gamma N(N-1)/2} \int {\cal D}\psi {\cal D}\xi ~ 
\exp \big( \sum_{i,j=0}^{\gamma (N-1)} \lambda^i \Xi_i w_{ij} 
\Psi_j \big) \nonumber \\
Z_V & = & \lambda^{-\gamma N(N-1)/2} \int {\cal D}\psi {\cal D}\xi ~ 
\exp \big( \sum_{i,j=0}^{\gamma (N-1)} \lambda^j \Xi_i w_{ij} 
\Psi_j \big) \\
Z_V \langle \Xi_i \Psi_j \rangle & = & \lambda^{-\gamma N(N-1)/2+i} 
\int {\cal D}\psi {\cal D}\xi ~ \Xi_i \Psi_j
\exp \big( \sum_{k,l=0}^{\gamma (N-1)} \lambda^k 
\Xi_k w_{kl} \Psi_l \big) \nonumber \\
Z_V \langle \Xi_i \Psi_j \rangle & = & \lambda^{-\gamma N(N-1)/2+j} 
\int {\cal D}\psi {\cal D}\xi ~ \Xi_i \Psi_j
\exp \big( \sum_{k,l=0}^{\gamma (N-1)} \lambda^l 
\Xi_k w_{kl} \Psi_l \big) 
\end{eqnarray}
\end{subequations}

The equality $\partial \ln Z_V / \partial \lambda \vert_{\lambda =1} = 0$
implies
\begin{subequations} \label{35}
\begin{eqnarray}
- {1\over 2} \gamma N (N-1) + \sum_{i,j=0}^{\gamma(N-1)} i~ w_{ij}
\langle \Xi_i \Psi_j \rangle & = & 0 \\
- {1\over 2} \gamma N (N-1) + \sum_{i,j=0}^{\gamma(N-1)} j~ w_{ij}
\langle \Xi_i \Psi_j \rangle & = & 0 
 \end{eqnarray}
\end{subequations}
On account of (\ref{30}), this is equivalent to a couple
of complex-conjugate equations
\begin{subequations} \label{36}
\begin{eqnarray}
{1\over 2} \gamma N(N-1) + N & = & \int {\rm d}^2 z ~ w(z,{\bar z})
\sum_{i,j=0}^{\gamma(N-1)} \langle \Xi_i \Psi_j \rangle
\partial^+ ( z^{i+1} {\bar z}^j ) \\
{1\over 2} \gamma N(N-1) + N & = & \int {\rm d}^2 z ~ w(z,{\bar z})
\sum_{i,j=0}^{\gamma(N-1)} \langle \Xi_i \Psi_j \rangle
\partial^- ( z^i {\bar z}^{j+1} )
\end{eqnarray}
\end{subequations}
where we have introduced derivative operators
\begin{equation} \label{37}
\partial^+ = {1\over 2} \left( {\partial \over \partial x} 
+ {1\over {\rm i}} {\partial \over \partial y} \right) 
\quad \quad \partial^- = {1\over 2} \left( {\partial \over \partial x} 
- {1\over {\rm i}} {\partial \over \partial y} \right) 
\end{equation}
$(\partial^+ \equiv \partial_z, \partial^- \equiv \partial_{{\bar z}})$.
They act on complex coordinates according to
$$\partial^+ z = 1, \quad \partial^+ {\bar z} = 0; \quad \quad
\partial^- z = 0, \quad \partial^- {\bar z} = 1.$$
With $\ln w(z,{\bar z}) = \Gamma n_0 \phi_b({\vek r})$, $\phi_b$
given by (\ref{17}), it is easy to verify the validity of equalities
\begin{equation} \label{38}
\int_V {\rm d}^2 z \left[ \partial^+ \ln w(z,{\bar z}) \right] z n_0
= \int_V {\rm d}^2 z \left[ \partial^- \ln w(z,{\bar z}) \right] {\bar z} n_0
= - {1\over 2} \gamma N^2
\end{equation}
After some algebra, Eqs.(\ref{36}) then take the form
\begin{subequations} 
\begin{eqnarray}
N\left( 1-{\gamma \over 2} \right) & = & \int_V {\rm d}^2 z ~
\partial^+ [z ~ n(z,{\bar z})] - \int_V {\rm d}^2 z ~ \left[ \partial^+
\ln w(z,{\bar z}) \right] ~ z ~ \delta n_V(z,{\bar z}) \label{39a} \\
N\left( 1-{\gamma \over 2} \right) & = & \int_V {\rm d}^2 z ~
\partial^- [{\bar z} ~ n(z,{\bar z})] - \int_V {\rm d}^2 z ~ \left[ \partial^-
\ln w(z,{\bar z}) \right] ~ {\bar z} ~ \delta n_V(z,{\bar z}) \label{39b}
\end{eqnarray}
\end{subequations}
with $\delta n_V(z,{\bar z}) = n_V(z,{\bar z}) - n_0$.
Let us denote by $\partial V$ the positively oriented contour 
enclosing domain $V$:
$\partial V$ is defined parametrically as follows $x = X(\phi), y = Y(\phi);
\phi_0 \le \phi \le \phi_1$.
In particular, the ellipse contour admits the parametrization 
$X(\phi) = a \cos \phi, Y = b \sin \phi; 0 \le \phi \le 2\pi$.
Integrals over $V$-domain can be expressed in terms of $\partial V$-contour
integrals according to formula
\begin{subequations} \label{40}
\begin{equation} \label{40a}
\int_V \left( {\partial Q \over \partial x} - {\partial P \over \partial y}
\right) {\rm d}x ~ {\rm d}y
= \int_{\partial V} \left( P ~ {\rm d}x + Q ~ {\rm d}y \right)
\end{equation}
where
\begin{eqnarray} \label{40b}
\int_{\partial V} P(x,y) {\rm d}x & = & \int_{\phi_0}^{\phi_1} {\rm d} \phi
~ P[X(\phi),Y(\phi)] X'(\phi) \nonumber \\
\int_{\partial V} Q(x,y) {\rm d}y & = & \int_{\phi_0}^{\phi_1} {\rm d} \phi
~ Q[X(\phi),Y(\phi)] Y'(\phi) 
\end{eqnarray}
\end{subequations} 
Thus, summing and subtracting Eqs.(\ref{39a}) and (\ref{39b}), 
one gets respectively
\begin{subequations} \label{41}
\begin{eqnarray}
\Gamma n_0 \int_V {\rm d}^2 r [{\vek r} \cdot {\vek E}_b({\vek r})]
\delta n_V({\vek r}) & = & N \left( 2 - {\Gamma \over 2} \right)
- \int_{\phi_0}^{\phi_1} {\rm d} \phi ~ n_V(X,Y) (X Y' - X' Y) 
\label{41a} \\ 
\Gamma n_0 \int_V {\rm d}^2 r [{\vek r} \times {\vek E}_b({\vek r})]_z
\delta n_V({\vek r}) & = & \int_{\phi_0}^{\phi_1} {\rm d} \phi ~ n_V(X,Y)
(X X' + Y Y') \label{41b} 
\end{eqnarray}
\end{subequations}
where $[{\vek r} \times {\vek E}_b]_z = x E_b^y - y E_b^x$.
Eq. (\ref{41a}) was obtained for the case of disk in Ref. \cite{Tellez}
[see Eq. (4.16)] and represents the generalization of the contact theorem: 
by removing the reference to the boundary and in the radius $R\to\infty$
limit, it results in the contact theorem for plane hard wall
\cite{Choquard3}, \cite{Totsuji}.
New complementary relation ({\ref{41b}) is informative for domain $V$ 
of generally deformed shape.

The equality $\partial [ Z_V \langle \Xi_i \Psi_j \rangle ] / \partial 
\lambda \vert_{\lambda =1} = 0$ results in
\begin{subequations} \label{42}
\begin{eqnarray}
\left[ - {1\over 2} \gamma N (N-1) + i \right] \langle \Xi_i \Psi_j \rangle
+ \sum_{k,l=0}^{\gamma(N-1)} k ~ w_{kl} \langle \Xi_i \Psi_j \Xi_k \Psi_l
\rangle & = & 0 \label{42a} \\ 
\left[ - {1\over 2} \gamma N (N-1) + j \right] \langle \Xi_i \Psi_j \rangle
+ \sum_{k,l=0}^{\gamma(N-1)} l ~ w_{kl} \langle \Xi_i \Psi_j \Xi_k \Psi_l
\rangle & = & 0 \label{42b} 
\end{eqnarray}
\end{subequations}
These relations can be rewritten with the aid of Eqs. (\ref{30}), (\ref{32})
and (\ref{35}) as follows
\begin{subequations} \label{43}
\begin{eqnarray}
(i+1) \langle \Xi_i \Psi_j \rangle & = & - \sum_{k,l=0}^{\gamma(N-1)}
(k+1) w_{kl} \langle \Xi_i \Psi_j \Xi_k \Psi_l \rangle^T \label{43a} \\
(j+1) \langle \Xi_i \Psi_j \rangle & = & - \sum_{k,l=0}^{\gamma(N-1)}
(l+1) w_{kl} \langle \Xi_i \Psi_j \Xi_k \Psi_l \rangle^T \label{43b}
\end{eqnarray}
\end{subequations}
It is a simple task to pass from (\ref{43}) to 
\begin{subequations} \label{44}
\begin{eqnarray}
w({\vek r}_1) \partial_1^+ \left[ {n_V({\vek r}_1) z_1 \over w({\vek r}_1)}
\right] & = & - \int_V {\rm d}^2 r_2 ~ w({\vek r}_2) \partial_2^+ \left[
{n_V^T({\vek r}_1,{\vek r}_2) z_2 \over w({\vek r}_2)} \right] \label{44a} \\
w({\vek r}_1) \partial_1^- \left[ {n_V({\vek r}_1) {\bar z}_1 
\over w({\vek r}_1)} \right] & = & 
- \int_V {\rm d}^2 r_2 ~ w({\vek r}_2) \partial_2^- \left[ {n_V^T({\vek r}_1,
{\vek r}_2) {\bar z}_2 \over w({\vek r}_2)} \right] \label{44b} 
\end{eqnarray}
\end{subequations}
with the obvious generalization of operators (\ref{37}):
$$\partial_i^+ z_j = \delta_{ij}, \quad \partial_i^+ {\bar z}_j = 0; \quad
\quad \partial_i^- z_j = 0, \quad \partial_i^- {\bar z}_j = \delta_{ij}.$$
Summing and subtracting (\ref{44a}) and (\ref{44b}) one finally arrives at
\begin{subequations} \label{45}
\begin{eqnarray}
\beta n_0 \int_V {\rm d}^2 r_2 ~ \left[ {\vek r}_2 \cdot {\vek E}_b({\vek r}_2)
\right] S_V({\vek r}_1 \vert {\vek r}_2) & = & - 2 n_V({\vek r}_1) - 
{\vek r}_1 \cdot \nabla_1 n_V({\vek r}_1) \nonumber \\
& & - \int_{\phi_0}^{\phi_1} {\rm d} \phi ~ n_V^T[{\vek r}_1;(X,Y)]
(XY' - X'Y) \label{45a} \\
\beta n_0 \int_V {\rm d}^2 r_2 \left[ {\vek r}_2 \times 
{\vek E}_b({\vek r}_2)\right]_z S_V({\vek r}_1 \vert {\vek r}_2) & = & 
- ({\vek r}_1 \times \nabla_1 )_z n_V({\vek r}_1) \nonumber \\
& & + \int_{\phi_0}^{\phi_1} {\rm d} \phi ~ n_V^T[{\vek r}_1;(X,Y)]
(XX' + YY') \label{45b}
\end{eqnarray}
\end{subequations}
Relation (\ref{45a}) with ${\vek r}_1 = {\vek 0}$ was derived for disk
in Ref. \cite{Tellez} [see Eq. (4.25)].
In the $R\to\infty$ limit of disk, it is related to the
dipole sum rule for plane hard wall \cite{Carnie}.
The complementary Eq. (\ref{45b}) is original.

\section{New sum rule}

Let us pose the following question: provided that the anticommuting
fields under consideration $\{ \xi_i^{(\alpha)} \}_{\alpha=1}^{\gamma}$
are mapped onto $\{ \xi_i^{(\alpha)}(t)\}_{\alpha=1}^{\gamma}$ by a
nearest-neighbor transformation
\begin{equation} \label{46}
{\partial \xi_i^{(\alpha)}(t) \over \partial t} = a_i \xi_{i+1}^{(\alpha)}(t)
+ b_i \xi_{i-1}^{(\alpha)}(t), \quad \quad \xi_i^{(\alpha)}(t=0) = 
\xi_i^{(\alpha)} \quad \quad (i=0,1,\ldots,N-1)
\end{equation}
with $a_{N-1} = b_0 = 0$ and $t$ being a free parameter, does there exist
a choice of the coefficients $\{ a_i, b_i \}$ for which also the composite
variables $\{ \Xi_i \}$ (\ref{22a}) transform according to a nearest-neighbor
scheme
\begin{equation} \label{47}
{\partial \Xi_i(t) \over \partial t} = {\tilde a}_i \Xi_{i+1}(t)
+ {\tilde b}_i \Xi_{i-1}(t), \quad \quad \Xi_i(t=0) = \Xi_i
\quad \quad [i=0,1,\ldots,\gamma(N-1)]
\end{equation}
with ${\tilde a}_{\gamma(N-1)} = {\tilde b}_0 = 0$?
The answer is affirmative \cite{Samaj2}: it can be proven by a direct 
computation that if one chooses
$$a_i = A(i+1) \quad \quad \quad b_i = B(N-i) \eqno(46')$$
the consequent $\{ \Xi_i(t) \}$ fulfil differential Eq. (\ref{47}) with
$${\tilde a}_i = A(i+1) \quad \quad \quad 
{\tilde b}_i = B[\gamma(N-1)+1-i] \eqno(47')$$
To determine Jacobian of the mapping, writting formally the solution
of (\ref{46}) as
\begin{equation} \label{48}
\xi_i^{(\alpha)}(t) = \sum_{j=0}^{N-1} c_{ij}(t) \xi_j^{(\alpha)}
\end{equation}
there holds
\begin{equation} \label{49}
{\partial c_{ij}(t) \over \partial t} = a_i c_{i+1,j}(t) 
+ b_i c_{i-1,j}(t), \quad \quad c_{ij}(0) = \delta_{ij}
\end{equation}
Jacobian equals to $\det c_{ij}(t) \vert_{i,j=0}^{N-1} \equiv 
\vert{\bf c}\vert$ for each of $\xi^{(\alpha)}$-components.
Its derivative with respect to $t$ is given by
\begin{equation} \label{50}
{\partial \over \partial t} \vert{\bf c}\vert = \sum_{i,j=0}^{N-1}
{\partial c_{ij} \over \partial t} C_{ij}
\end{equation}
where $C_{ij}(t)$ is the cofactor of element $c_{ij}(t)$.
In combination with Eq. (\ref{49}), the orthogonality condition
\begin{equation} \label{51}
\sum_{j=0}^{N-1} c_{kj} C_{ij} = \delta_{ik} \vert{\bf c}\vert
\end{equation}
thus leads to $\partial \vert{\bf c}\vert
/\partial t = 0$, i.e., $\vert{\bf c}\vert = {\rm const} = 1$ for
each $\xi^{(\alpha)}$-component.
We conclude that Jacobian $=1$.
For our purpose it is sufficient to consider transformation (\ref{46}),
(\ref{47}) with $A=1$ and $B=0$; the explicit solution reads
\begin{equation} \label{52}
\xi_i^{(\alpha)}(t) = \sum_{j=i}^{N-1} {j\choose i} t^{j-i} \xi_j^{(\alpha)},
\quad \quad
\Xi_i(t) = \sum_{j=i}^{\gamma(N-1)} {j\choose i} t^{j-i} \Xi_j
\end{equation}

The insertion of transformation (\ref{52}) into the partition function,
\begin{eqnarray} \label{53}
Z_V & = & \int {\cal D} \psi {\cal D} \xi(t) 
\exp \big\{ \sum_{i,j=0}^{\gamma(N-1)}
\Xi_i(t) w_{ij} \Psi_j \big\} \nonumber \\
& = & \int {\cal D}\psi {\cal D}\xi ~ \exp \big\{ \sum_{i,j=0}^{\gamma(N-1)}
\left[ \Xi_i + t (i+1) \Xi_{i+1} + O(t^2) \right] w_{ij} \Psi_j \big\}
\end{eqnarray}
with $\Xi_{\gamma(N-1)+1} \equiv 0$ automatically assumed, and the 
consequent application of condition
$\partial \ln Z_V / \partial t \vert_{t=0} = 0$ lead to
\begin{subequations} \label{54}
\begin{eqnarray}
\sum_{i,j=0}^{\gamma (N-1)} (i+1) w_{ij} \langle \Xi_{i+1} \Psi_j \rangle
& = & 0  \label{54a} \\
\sum_{i,j=0}^{\gamma (N-1)} (j+1) w_{ij} \langle \Xi_i \Psi_{j+1} \rangle
& = & 0  \label{54b}
\end{eqnarray}
\end{subequations}
where the second formula originates from the $t$-transformation of
$\{ \psi^{(\alpha)} \}$ anticommuting fields.
The equivalent couple of complex-conjugate equations
\begin{subequations} \label{55}
\begin{eqnarray}
\int_V {\rm d}^2 r ~ [ \partial^+ \ln w({\vek r}) ] n_V({\vek r}) & = &
\int_V {\rm d}^2 r ~ \partial^+ n_V({\vek r})  \label{55a} \\
\int_V {\rm d}^2 r ~ [ \partial^- \ln w({\vek r}) ] n_V({\vek r}) & = &
\int_V {\rm d}^2 r ~ \partial^- n_V({\vek r})  \label{55b}
\end{eqnarray}
\end{subequations}
is expressible by using the previously developed formalism as follows
\begin{subequations} \label{56}
\begin{eqnarray}
\Gamma n_0 \int_V {\rm d}^2 r ~ E_b^x({\vek r}) n_V({\vek r}) & = &
- \int_{\phi_0}^{\phi_1} {\rm d} \phi ~ n_V(X,Y) Y' \label{56a} \\
\Gamma n_0 \int_V {\rm d}^2 r ~ E_b^y({\vek r}) n_V({\vek r}) & = &
\int_{\phi_0}^{\phi_1} {\rm d} \phi ~ n_V(X,Y) X' \label{56b}
\end{eqnarray}
\end{subequations}

The $t$-independence of
\begin{eqnarray} \label{57}
Z_V \langle \Xi_i(t) \Psi_j \rangle_t & = & \int {\cal D} \psi {\cal D} \xi
\left[ \Xi_i + t (i+1) \Xi_{i+1} + O(t^2) \right] \Psi_j  \nonumber \\
& & \times \exp \big\{ \sum_{k,l=0}^{\gamma(N-1)} \left[ \Xi_k + 
t(k+1) \Xi_{k+1} + O(t^2) \right] w_{kl} \Psi_l \big\}
\end{eqnarray}
and similarly of $Z_V \langle \Xi_i \Psi_j(t)\rangle_t$ 
manifests itself at the lowest $t^1$ level as
\begin{subequations} \label{58}
\begin{eqnarray}
(i+1) \langle \Xi_{i+1} \Psi_j \rangle + \sum_{k,l=0}^{\gamma(N-1)}
(k+1) w_{kl} \langle \Xi_i \Psi_j \Xi_{k+1} \Psi_l \rangle & = & 0 
\label{58a} \\
(j+1) \langle \Xi_i \Psi_{j+1} \rangle + \sum_{k,l=0}^{\gamma(N-1)}
(l+1) w_{kl} \langle \Xi_i \Psi_j \Xi_k \Psi_{l+1} \rangle & = & 0 
\label{58b}
\end{eqnarray}
\end{subequations}
Due to (\ref{54}), the four-correlators $\langle \Xi_i \Psi_j \Xi_{k+1}
\Psi_l \rangle$ in (\ref{58a}) and $\langle \Xi_i \Psi_j \Xi_k \Psi_{l+1} 
\rangle$ in (\ref{58b}) can be substituted by the truncated ones
$\langle \Xi_i \Psi_j \Xi_{k+1} \Psi_l \rangle^T$ and 
$\langle \Xi_i \Psi_j \Xi_k \Psi_{l+1} \rangle^T$, respectively.
Eqs. (\ref{58a}) and (\ref{58b}) are thus expressible as
\begin{subequations} \label{59}
\begin{eqnarray}
w({\vek r}_1) \partial^+_1 \left[ {n_V({\vek r}_1) \over w({\vek r}_1)} \right]
& = & - \int_V {\rm d}^2 r_2 ~ w({\vek r}_2) \partial_2^+ \left[
n_V^T({\vek r}_1,{\vek r}_2) \over w({\vek r}_2) \right] \label{59a} \\
w({\vek r}_1) \partial^-_1 \left[ {n_V({\vek r}_1) \over w({\vek r}_1)} \right]
& = & - \int_V {\rm d}^2 r_2 ~ w({\vek r}_2) \partial_2^- \left[
n_V^T({\vek r}_1,{\vek r}_2) \over w({\vek r}_2) \right] \label{59b}
\end{eqnarray}
\end{subequations}
These relations can be further simplified to the form
\begin{subequations} \label{60}
\begin{eqnarray}
\int_V {\rm d}^2 r_2 \left[ \partial_2^+ \ln w({\vek r}_2) \right]
S_V({\vek r}_1 \vert {\vek r}_2) / q^2 & = & \partial_1^+ n_V({\vek r}_1)
+ \int_V {\rm d}^2 r_2 ~ \partial_2^+ n_V^T({\vek r}_1,{\vek r}_2) 
\label{60a} \\
\int_V {\rm d}^2 r_2 \left[ \partial_2^- \ln w({\vek r}_2) \right]
S_V({\vek r}_1 \vert {\vek r}_2) / q^2 & = & \partial_1^- n_V({\vek r}_1)
+ \int_V {\rm d}^2 r_2 ~ \partial_2^- n_V^T({\vek r}_1,{\vek r}_2) \label{60b}
\end{eqnarray}
\end{subequations}
For the elliptic $V$-domain of interest it holds
$\ln w({\vek r}) = \Gamma n_0 ( {\rm const} - \pi T_V^x x^2 - \pi T_V^y y^2)$.
Summing and subtracting Eqs. ({\ref 60a}) and ({\ref 60b}) one then finds
respectively
\begin{subequations} \label{61}
\begin{eqnarray}
- 2 \pi \beta n_0 T_V^x \int_V {\rm d}^2 r_2 ~ x_2 S_V({\vek r}_1 \vert 
{\vek r}_2) & = & {\partial \over \partial x_1} n_V({\vek r}_1) +
\int_V {\rm d}^2 r_2 {\partial \over \partial x_2} 
n_V^T({\vek r}_1,{\vek r}_2)  \label{61a} \\
- 2 \pi \beta n_0 T_V^y \int_V {\rm d}^2 r_2 ~ y_2 S_V({\vek r}_1 \vert 
{\vek r}_2) & = & {\partial \over \partial y_1} n_V({\vek r}_1) +
\int_V {\rm d}^2 r_2 {\partial \over \partial y_2} 
n_V^T({\vek r}_1,{\vek r}_2)  \label{61b}
\end{eqnarray}
\end{subequations}

To get the diagonal elements of the dielectric susceptibility tensor
$\chi_V$ (\ref{12}), one applies $\int_V {\rm d}^2 r_1 ~ x_1$ to (\ref{61a})
and $\int_V {\rm d}^2 r_1 ~ y_1$ to (\ref{61b}), with the result
\begin{subequations} \label{62}
\begin{eqnarray}
\chi_V^x & = & {1\over 2\pi T_V^x} - {a+b\over ab}\ {1\over 2\pi^2 n_0
q^2} \int_0^{2\pi} {\rm d}\phi \int_V {\rm d}^2 r_1 ~
x_1 S_V[{\vek r}_1 \vert (X,Y)] \cos \phi   \label{62a} \\
\chi_V^y & = & {1\over 2\pi T_V^y} - {a+b\over ab}\ {1\over 2\pi^2 n_0
q^2} \int_0^{2\pi} {\rm d}\phi \int_V {\rm d}^2 r_1 ~
y_1 S_V[{\vek r}_1 \vert (X,Y)] \sin \phi   \label{62b}
\end{eqnarray}
\end{subequations}
Here, integration per partes was combined with the sum rule (\ref{31})
to obtain
$$\int_V {\rm d}^2 r_1 ~ x_1 {\partial \over \partial x_1} n_V({\vek r}_1)
= \int_V {\rm d}^2 r_1 {\partial \over \partial x_1} 
\left[ x_1 n_V({\vek r}_1) \right] - n_0 \vert V \vert,$$
$\vert V \vert = \pi a b$, and analogously for the $y$-component.
Eqs. (\ref{62}) become simpler in the symmetric case of disk of radius $R$,
$\chi^x_V = \chi^y_V = {\bar\chi}_V$,
\begin{eqnarray} \label{63}
{\bar\chi}_V & = & {1\over \pi} - {1\over R} \ {1\over \pi n_0 q^2}
\int_0^{2\pi} {{\rm d}\phi \over 2\pi} \int_V {\rm d}^2 r_1
S_V[(r_1,\phi_1)\vert (R,\phi)] r_1 \cos(\phi_1-\phi) \nonumber \\
& = & {1\over \pi} - {1\over R} \ {1\over \pi n_0 q^2} \int_V
{\rm d}^2 r ~ x ~ S_V[{\vek r}\vert (R,0)] \\
& = & {1\over \pi} - {1\over R} \ {1\over \pi n_0 q^2} \int_V
{\rm d}^2 r ~ y ~ S_V[{\vek r}\vert (0,R)] \nonumber
\end{eqnarray}
where the dependence of $S_V[(r_1,\phi_1)\vert (R,\phi)]$ on the angle
difference $\vert \phi_1 - \phi \vert$ was taken into account.
Relation (\ref{63}) can be formally rewritten as
\begin{eqnarray} \label{64}
{\bar\chi}_V & = & {1\over \pi} - {1\over R} \ {1\over \pi n_0 q^2} 
\langle P^x \hat \rho[(R,0)] \rangle_V^T \nonumber \\
& = & {1\over \pi} - {1\over R} \ {1\over \pi n_0 q^2} 
\langle P^y \hat \rho[(0,R)] \rangle_V^T
\end{eqnarray}
The final results (\ref{62})--(\ref{64}) mean an explicit split of
$\chi_V^i$ into its asymptotic $1/(2\pi T_V^i)$ part 
[see the prediction (\ref{15}) of macroscopic electrostatics]
and the finite-size correction term.
To show this fact for disk, with regard to the sum rule (\ref{10}) 
one can write
\begin{equation} \label{65}
\int_V {\rm d}^2 r ~ x S_V[{\vek r} \vert (R,0)] = 
- \int_V {\rm d}^2 r ~ (R-x) S_V[{\vek r} \vert (R,0)]
\end{equation}
Moving the reference to the boundary, $x' = R-x$ and $y'=y$, the integrals
on the rhs of (\ref{63}) reflect the dipole moment seen from the boundary,
which is known to converge to a finite value in the thermodynamic limit.
We therefore conclude that the correction term $\sim 1/R$.
Owing to a slow power-law decay of correlations along a plane wall
\cite{Jancovici5},\cite{Jancovici6}, 
one has to be cautious when identifying the integrals
of type (\ref{65}) with their asymptotic hard-wall counterparts.
Possible vagaries and a check of Eq. (\ref{63}) are documented
in Appendix via the exactly solvable 2d OCP at coupling $\Gamma =2$.

In conclusion, although the above results (\ref{62})--(\ref{64}) were
derived strictly for coupling constant $\Gamma = 2*$positive integer, 
it is reasonable to suppose their validity in the whole fluid regime.
The extension of the treatment to the case of the charged wall and in
the presence of image forces is straightforward.
Potential generalization of the results to higher dimensions and to 
the TCP requires, due to the lack of the fermionic formalism,
to search for a new method which, on the one hand, reproduces our findings 
for the 2d OCP and, on the other hand, admits a more general applicability
like the linear-response theory.  

\section*{Acknowledgment}
I am very much indebted to Bernard Jancovici for careful reading of 
the manuscript and for useful comments.
This work was supported by grant VEGA no 2/7174/20.

\renewcommand{\theequation}{A\arabic{equation}}
\setcounter{equation}{0}

\section*{Appendix}
When $\Gamma =2$ $(\gamma =1)$, the 2d OCP on disk of radius $R$
is exactly solvable \cite{Jancovici3},\cite{Jancovici4},\cite{Jancovici5}.
The fermionic correlators are given by
\begin{equation} \label{A1}
\langle \Xi_i \Psi_j \rangle = {1\over w_i} \delta_{ij}, \quad \quad
\langle \Xi_i \Psi_j \Xi_k \Psi_l \rangle = {1\over w_i w_k} 
\left( \delta_{ij} \delta_{kl} - \delta_{il} \delta_{jk} \right)
\end{equation}
etc., where $\{ w_i \}$ (\ref{27}) are diagonal interaction strengths,
\begin{equation} \label{A2}
w_i = \int_V {\rm d}^2 r ~ r^{2i} w(r) = \pi \int_0^N {\rm d}t ~
t^i \exp(-t)
\end{equation}
written in the units of $\pi n_0 = 1$.
The dielectric susceptibility tensor is expressible as
\begin{equation} \label{A3}
{\bar\chi}_V = {1\over \pi R^2} {\rm Re} \left\{ \int_V {\rm d}^2 r ~ 
r^2 n(r) + \int_V {\rm d}^2 r_1 \int_V {\rm d}^2 r_2 ~ {\bar z}_1 z_2
n_V^T({\vek r}_1,{\vek r}_2) \right\}
\end{equation}
It is straightforward to verify the validity of relations
\begin{subequations} \label{A4}
\begin{eqnarray}
\int_V {\rm d}^2 r ~ r^2 n(r) & = & \sum_{i=0}^{N-1} {w_{i+1} \over w_i} \\
\int_V {\rm d}^2 r_1 \int_V {\rm d}^2 r_2 ~ {\bar z}_1 z_2
n_V^T({\vek r}_1,{\vek r}_2) & = & - \sum_{i=1}^{N-1} {w_i \over w_{i-1}} 
\end{eqnarray}
\end{subequations}
so that ${\bar\chi}_V = w_N/(\pi N w_{N-1})$.
By integration per partes one derives
$w_i = i w_{i-1} - \pi N^i \exp(-N)$. 
Consequently,
\begin{equation} \label{A5}
{\bar\chi}_V = {1\over \pi} - {N^{N-1} \exp(-N) \over w_{N-1}}
\end{equation}

On the other hand,
\begin{equation} \label{A6}
{1\over q^2} \int_V {\rm d}^2 r ~ x ~ S_V[{\vek r}\vert(R,0)] = R ~ n(R)
+ \int_V {\rm d}^2 r ~ x ~ n_V^T[{\vek r};(R,0)]
\end{equation}
Since
\begin{subequations} \label{A7}
\begin{eqnarray}
R ~ n(R) & = & w(R) \sum_{i=0}^{N-1} {R^{2i+1} \over w_i} \\
\int_V {\rm d}^2 r ~ x ~ n_V^T[{\vek r};(R,0)] & = & - w(R)
\sum_{i=1}^{N-1} {R^{2i-1} \over w_{i-1}}
\end{eqnarray}
\end{subequations}
one arrives at
\begin{equation} \label{A8}
{1\over q^2} \int_V {\rm d}^2 r ~ x ~ S_V[{\vek r}\vert (R,0)] =
{N^N \exp(-N) \over R w_{N-1}}
\end{equation}
Inserting (\ref{A8}) into (\ref{63}), the exact result (\ref{A5})
is recovered (in the units of $\pi n_0 = 1$).

As $N\to\infty$, the asymptotic form of $w_{N-1}$ (\ref{A2}) can be 
calculated by the saddle-point method in the gaussian approximation:
\begin{equation} \label{A9}
w_{N-1} \sim {\pi^{3/2} \over \sqrt{2}} R N^{N-1} \exp(-N)
\left[ 1 + O\left( {1\over R} \right) \right]
\end{equation}
Consequently, the quantity (\ref{A8}), transcribed according to (\ref{65}),
acquires the finite value
\begin{equation} \label{A10}
\lim_{N\to\infty} - {1\over q^2} \int_V {\rm d}^2 r ~ (R-x) 
S_V[{\vek r}\vert (R,0)] = {\sqrt{2} \over \pi^{3/2}}
\end{equation}
as was expected.

One may be tempted to identify formula (\ref{A10}) with its obvious plane
hard-wall counterpart.
Using the explicit result \cite{Jancovici5} for the hard wall localized
at $x=0$ (plasma appears in the half-space $x\ge 0$),
\begin{subequations} \label{A11}
\begin{eqnarray}
n(x) & = & n_0 {2\over \sqrt{\pi}} \int_0^{\infty} {\exp\left[ -
(t-x\sqrt{2})^2 \right] \over 1 + \phi(t)} {\rm d}t \\
n^T(x_1,x_2;\vert y_1 - y_2 \vert) & = & - n_0^2 \exp\left[ -
(x_1-x_2)^2 \right] \\
& \times & \left\vert {2\over \sqrt{\pi}} \int_0^{\infty} {\exp\left\{
-\left[ t-(x_1+x_2)/\sqrt{2} \right]^2 - {\rm i}t(y_1-y_2)\sqrt{2} 
\right\} \over 1+\phi(t)} {\rm d}t \right\vert^2 \nonumber
\end{eqnarray}
\end{subequations}
where $\phi$ is the error function
$\phi(t) = (2/ \sqrt{\pi}) \int_0^t \exp(-u^2) {\rm d}u$,
one obtains
\begin{equation} \label{A12}
- {1\over q^2} \int_0^{\infty} {\rm d}x ~ x \int_{-\infty}^{\infty} 
{\rm d}y ~ S(0,x;y) = {1\over \sqrt{2} \pi^{3/2}}
\end{equation}
which differs from (\ref{A10}) by factor 2.
This discrepancy is intuitively associated with the slow power-law
decay of correlations along the plane wall: an arbitrarily small
deformation of the boundary towards the circle has a strong impact on
this property.
The analytical (Debye-H\"uckel approximation) and numerical (Monte-Carlo
simulation) studies of the surface charge correlations for finite
Coulomb systems were given in Ref. \cite{Choquard4}.
The asymptotic form of these correlations is expected to depend on 
the shape of the plasma, but to be otherwise universal.
The exact $\Gamma=2$ solution for the ``soft-wall'' version of the
2d OCP with a quadrupolar field \cite{Forrester1}, corresponding
to a very large elliptic background, supports this finding.


\newpage

\begin{thebibliography}{9}

\bibitem{Martin} Ph.A. Martin, 
{\it Rev. Mod. Phys.} {\bf 60}:1075 (1988). 

\bibitem{Stillinger1} F.H. Stillinger and R. Lovett, 
{\it J. Chem. Phys.} {\bf 48}:3858 (1968). 

\bibitem{Stillinger2} F.H. Stillinger and R. Lovett, 
{\it J. Chem. Phys.} {\bf 49}:1991 (1968). 

\bibitem{Pines} D. Pines and Ph. Nozi\` eres, {\it The Theory of Quantum
Liquids} (Benjamin, New York, 1966).

\bibitem{Vieillefosse1} P. Vieillefosse and J.P. Hansen, 
{\it Phys. Rev. A} {\bf 12}:1106 (1975). 

\bibitem{Baus} M. Baus, 
{\it J. Phys. A: Math. Gen.} {\bf 11}:2451 (1978). 

\bibitem{Hauge} E.H. Hauge and P.C. Hemmer, 
{\it Phys. Norvegica} {\bf 5}:209 (1981). 

\bibitem{Vieillefosse2} P. Vieillefosse, 
{\it J. Stat. Phys.} {\bf 41}:1015 (1985). 

\bibitem{Suttorp} L.G. Suttorp and J.S. Cohen,
{\it Physica A} {\bf 133}:357 (1985). 

\bibitem{Kalinay} P. Kalinay, P. Marko\v s, L. \v Samaj, 
and I. Trav\v enec, 
The Sixth-Moment Sum Rule for the Pair Correlations of the 2d OCP:
Exact Result, cond-mat/9907024, 
{\it J. Stat. Phys.} {\bf 98} (2000) Issue 3/4.

\bibitem{Deutsch} C. Deutsch and M. Lavaud,
{\it Phys. Rev. A} {\bf 9}:2598 (1974). 

\bibitem{Jancovici1} B. Jancovici,
A Sum Rule for the 2d TCP, LPT Orsay 99-59, cond-mat/9907365,
accepted for publication in {\it J. Stat. Phys.} 

\bibitem{Jancovici2} B. Jancovici, P. Kalinay, and L. \v Samaj,
Another Derivation of a Sum Rule for the 2d TCP, cond-mat/9909261,
accepted for publication in {\it Physica A}.

\bibitem{Choquard1} Ph. Choquard, B. Piller, and R. Rentsch,
{\it J. Stat. Phys.} {\bf 43}:197 (1985). 

\bibitem{Choquard2} Ph. Choquard, B. Piller, and R. Rentsch,
{\it J. Stat. Phys.} {\bf 46}:599 (1986). 
 
\bibitem{Jancovici3} B. Jancovici,
{\it Phys. Rev. Lett.} {\bf 46}:386 (1981). 

\bibitem{Jancovici4} B. Jancovici, {\it Inhomogeneous Fluids},
D. Henderson, ed. (Dekker, New York, 1992), pp.201-237.

\bibitem{Samaj1} L. \v Samaj and J.K. Percus,
{\it J. Stat. Phys.} {\bf 80}:811 (1995). 

\bibitem{Samaj2} L. \v Samaj, P. Kalinay and I. Trav\v enec,
{\it J. Phys. A: Math. Gen.} {\bf 31}:4149 (1998). 

\bibitem{Tellez} G. T\'ellez and P.J. Forrester,
{\it J. Stat. Phys.} {\bf 97}:489 (1999).

\bibitem{Berezin} F.A. Berezin,
{\it The Method of Second Quantization} (New York, Academic Press, 1966).

\bibitem{Choquard3} Ph. Choquard, P. Favre, and Ch. Gruber,
{\it J. Stat. Phys.} {\bf 23}:405 (1980). 

\bibitem{Totsuji} H. Totsuji, 
{\it J. Chem. Phys.} {\bf 75}:871 (1981). 

\bibitem{Carnie} S.L. Carnie,
{\it J. Chem. Phys.} {\bf 78}:2742 (1983). 

\bibitem{Jancovici5} B. Jancovici, 
{\it J. Stat. Phys.} {\bf 28}:43 (1982). 

\bibitem{Jancovici6} B. Jancovici, 
{\it J. Stat. Phys.} {\bf 29}:263 (1982). 

\bibitem{Choquard4} Ph. Choquard, B. Piller, R. Rentsch, and P. Vieillefosse,
{\it J. Stat. Phys.} {\bf 55}:1185 (1989).

\bibitem{Forrester1} P.J. Forrester and B. Jancovici,
{\it Int. J. Mod. Phys. A} {\bf 11}:941 (1996).


\end{thebibliography}
\end{document}